\documentclass[twocolumn,twoside,slac_two]{revtex4}
\usepackage{graphicx}
\usepackage{fancyhdr}
\def\mapgeq{\mathbin{\lower.3ex\hbox{$\buildrel>\over{\smash{\scriptstyle\sim}\vphantom{_x}}$}}}
\def\mapleq{\mathbin{\lower.3ex\hbox{$\buildrel<\over{\smash{\scriptstyle\sim}\vphantom{_x}}$}}}
\def\mapgeqeq{\mathbi{\lower.3ex\hbox{$\buildrel>\over{\smash{\scriptstyle\approx}\vphantom{_2}}$}}}
\def\mapleqeq{\mathbin{\lower.3ex\hbox{$\buildrel<\over{\smash{\scriptstyle\approx}\vphantom{_2}}$}}}
\def\Journal#1#2#3#4{{#1} {\bf #2} (#4) #3}
\def\NPB{Nucl. Phys. B}

\def\PLB{{Phys. Lett.} B}

\def\PRL{Phys. Rev. Lett.}

\def\PRD{Phys. Rev. D}

\def\PTP{Prog. Theor. Phys.}
\def\JHEP{JHEP}

\def\JETPUSSR{Sov. Phys. JETP}

\def\ZETP{Zh. Eksp. Teor. Fiz.}

\def\IJMPE{Int. J. Mod. Phys. E}

\def\Erratum{Erratum-ibid}

\begin{document}

\title{Approximately $\mu$-$\tau$ Symmetric Minimal Seesaw Mechanism and Leptonic CP Violation} 

\author{Teppei Baba}
\affiliation{Department of Physics, Tokai University, 1117 Kitakaname, Hiratsuka, Kanagawa 259-1292, Japan}
%

%%%%%%%%%%%%%%%%%%%%%%%%%%%%%%%
%%%%%%%%%%%%%%%%%%%%%%%%%%%%%%%
%%%%%%%%%%%%%%%%%%%%%%%%%%%%%%%
\begin{abstract}
We estimate the effect of leptonic CP violating phases in the minimal seesaw model 
with two heavy neutrinos $N$ by imposing the approximately $\mu-\tau$ symmetry on the model.
For $N$ subject to the $\mu-\tau$ symmetry, we find that neutrinos show the normal 
mass hierarchy and obtain the general phase structure that contains only two phases which arise from $\mu-\tau$
 symmetry breaking terms. To perform similar consideration for $N$ blind to the $\mu-\tau$ symmetry necessarily 
giving the inverted mass hierarchy, we assume the same phase structure.
Our phases consist of six phases that need to take care of generic phases of 
flavor neutrino masses, from which one can obtain one Dirac phase and one Majorana phase as observable 
phases (because one neutrino is massless). The Dirac CP violating phase $\delta_{CP}(=\delta+\rho)$ depends on the 
$\nu_e-\nu_{\tau}$ ($\nu_e-\nu_{\mu}$) mixing phase $\delta$ ($\rho$). The Majorana CP violating 
phase is suppressed because it depend on $\rho$, which is found to 
be associated with almost real quantity.
\end{abstract}
%%%%%%%%%%%%%%%%%%%%%%%%%%%%%%%
%%%%%%%%%%%%%%%%%%%%%%%%%%%%%%%
%%%%%%%%%%%%%%%%%%%%%%%%%%%%%%%

\maketitle

\thispagestyle{fancy}

\section{Introduction}
The $\mu-\tau$ symmetry is the symmetry that provides the consistent 
sizes of mixing angles with those indicated by experimental data \cite{mu-tau}. 
In the minimal seesaw model \cite{Seesaw model} with two heavy right-handed neutrinos \cite{Minimal seesaw model},
 we obtain
that 1) the normal mass hierarchy is realized if N is subject to the $\mu-\tau$ 
symmetry while the inverted mass hierarchy is realized if N is blind to the 
$\mu-\tau$ symmetry;
2) we understand how CP violation depends on phases of flavor neutrinos masses;
3) Dirac CP violating phase depends on $\nu_e-\nu_{\tau}$ mixing phase $\delta$ 
and Majorana CP violation depends on 
$\nu_e-\nu_{\mu}$ phase $\rho$, which turns out to be suppressed.

The mass term for neutrinos in the minimal seesaw model is defined as follows:
%%%%%%%%%%%%%%%%%%%%
\begin{equation}
L_{mass}  = - \overline {e_R } Y_l LH_d   -  \overline {N} Y_{\nu}  LH_u  - \frac
{1}{2}N^T M_R N,
\label{L_mass}
\end{equation}
%%%%%%%%%%%%%%%%%%%%
where $Y_l$ and $Y_{\nu}$ are Yukawa couplings, and $M_R$ is a mass mtrix of $N$.
Three flavor leptons are denoted by $L$ as SU(2)$_L$-doublets and by $e_R$ as SU
(2)$_L$-singlets, $H_{u,d}$ denote two Higgses
and two flavor heavy neutrinos are denoted by $N$ as SU(2)$_L$ singlets.
If there are three flavorfor $N$, $N$ can be any combination of $N_{e,\mu,\tau}$.
We use $e,\mu,\tau$ as the suffix of $N$.
We have chosen the base in which $Y_l$ is diagonalized.
The coupling $Y_{\nu}$ and the mass matrix $M_R$ are parametraized by
%%%%%%%%%%%%%%%%%%%%
\begin{equation}
 Y_\nu   = \left( {\begin{array}{*{20}c}
   {h_{ f_1  e} } & {h_{ f_1  \mu } } & {h_{ f_1 \tau } }  \\
   {h_{ f_2  e} } & {h_{ f_2  \mu } } & {h_{ f_2  \tau } }  \\
\end{array}} \right),\,\,\,
M_R  = \left( {\begin{array}{*{20}c}
   {M_{Rf_1 f_1 } } & {M_{Rf_1 f_2 } }  \\
   {M_{Rf_2 f_1 } } & {M_{Rf_2 f_2 } }  \\
\end{array}} \right).
\label{Definition of Y_nu,N and M_R}
\end{equation}
%%%%%%%%%%%%%%%%%%%%
where $f_1\left(  \ne f_2  \right) =e,\mu,\tau$ or other possible combinations 
surviving as two ``light" heavy neutrinos.

The seesaw mechanism yields a neutrino mass matrix:
%%%%%%%%%%%%%%%%%%%%
\begin{equation}
M_\nu   =  - \left\langle {H_u } \right\rangle ^2 Y_\nu  ^T M_R^{ - 1} Y_\nu   = 
\left( {\begin{array}{*{20}c}
   {M_{ee} } & {M_{e\mu } } & {M_{e\tau } }  \\
   {M_{e\mu } } & {M_{\mu \mu } } & {M_{\mu \tau } }  \\
   {M_{e\tau } } & {M_{\mu \tau } } & {M_{\tau \tau } }  \\
\end{array}} \right),
\label{M_nu}
\end{equation}
%%%%%%%%%%%%%%%%%%%%
which satisfies $\det\left( M_{\nu} \right) =0$ in the minimal seesaw model.
Hence, at least, one of neutrinos does not have a mass.
$M_\nu$ is diagonalized by Pontecorvo-Maki-Nakagawa-Sakata Matirix $U_{PMNS}=U_{\nu}K$
 \cite{PMNS}.
We parametrize $U_{\nu}$ and $K$ to be \cite{Dependance of CP phase on X and Y}:
%%%%%%%%%%%%%%%%%%%%
\begin{eqnarray}
&&
{\scriptsize
\begin{array}{l}
 U_\nu   = \left( {\begin{array}{*{20}c}
   1 & 0 & 0  \\
   0 & {e^{i\gamma } } & 0  \\
   0 & 0 & {e^{ - i\gamma } }  \\
\end{array}} \right) \\ 
 {\rm{      }}\left( {\begin{array}{*{20}c}
   {c_{13} c_{12} } & {c_{13} s_{12} e^{i\rho } } & {s_{13} e^{ - i\delta } }  \\
   {\left( \begin{array}{l}
  - s_{12} c_{23} e^{ - i\rho }  \\ 
  - s_{13} c_{12} s_{23} e^{i\delta }  \\ 
 \end{array} \right)} & {\left( \begin{array}{l}
 c_{12} c_{23}  \\ 
  - s_{12} s_{13} s_{23} e^{i\left( {\delta  + \rho } \right)}  \\ 
 \end{array} \right)} & {c_{13} s_{23} }  \\
   {\left( \begin{array}{l}
 s_{12} e^{ - i\rho } s_{23}  \\ 
  - s_{13} c_{12} c_{23} e^{i\delta }  \\ 
 \end{array} \right)} & {\left( \begin{array}{l}
  - c_{12} s_{23}  \\ 
  - s_{12} s_{13} c_{23} e^{i\left( {\delta  + \rho } \right)}  \\ 
 \end{array} \right)} & {c_{13} c_{23} }  \\
\end{array}} \right) \\ 
 \end{array}
}, 
\nonumber\\
&&K = diag\left( {e^{i \beta _1 } ,e^{i\beta _2 } ,e^{i\beta _3 } } \right),
\label{U_PMNS}
\end{eqnarray}
%%%%%%%%%%%%%%%%%%%%
where $U_{PMNS}$ has three neurino mixing angles $\theta_{12},\theta_{13}$ and 
$\theta_{23}$ and six leptonic CP violating phases 
that are defined as Dirac CP phases: $\delta$, $\rho$ and $\gamma$ and Majorana 
CP phases: $\beta_{1},\beta_{2}$ and $\beta_{3}$ \cite{CP Violation},
$c_{ij}$ and $s_{ij}$ represent $\cos\theta_{ij}$ and $\sin\theta_{ij}$.
The PMNS unitary matrix converts the left-handed flavor neutrinos into massive 
neutrinos as $\nu _f  = \sum\limits_j {\left( {U_{PMNS} } \right)_{fj} \nu _j }$,
(f = e,$\mu ,\tau$ i = 1,2,3). Note that $\gamma$ is the $\mu-\tau$ symmetery 
breaking quantity which is, therefore, generically small.

The phases $\rho$, $\gamma$ and one of Majorana phases are redundant parameters 
and therefore the PDG (particle data group) version \cite{PDG} of the PMNS matrix 
$U_{PMNS}^{PDG}=U_{\nu}^{PDG}{K'}$ can be
 defined by removing these redundant phases from $U_{PMNS}$ to be:
%%%%%%%%%%%%%%%%%%%%
\begin{eqnarray}
&&{\scriptsize
\begin{array}{l}
 U_\nu ^{PDG}  \\ 
  = \left( {\begin{array}{*{20}c}
   {c_{13} c_{12} } & {c_{13} s_{12} } & {s_{13} e^{ - i\delta _{CP} } }  \\
   {\left( \begin{array}{l}
  - s_{12} c_{23}  \\ 
  - s_{13} c_{12} s_{23}  \\ 
 {\rm{ }} \cdot e^{i\delta _{CP} }  \\ 
 \end{array} \right)} & {\left( \begin{array}{l}
 c_{12} c_{23}  \\ 
  - s_{12} s_{13} s_{23}  \\ 
 {\rm{ }} \cdot e^{i\delta _{CP} }  \\ 
 \end{array} \right)} & {c_{13} s_{23} }  \\
   {\left( \begin{array}{l}
 s_{12} s_{23}  \\ 
  - s_{13} c_{12} c_{23}  \\ 
 {\rm{ }} \cdot e^{i\delta _{CP} }  \\ 
 \end{array} \right)} & {\left( \begin{array}{l}
  - c_{12} s_{23}  \\ 
  - s_{12} s_{13} c_{23}  \\ 
 {\rm{ }} \cdot e^{i\delta _{CP} }  \\ 
 \end{array} \right)} & {c_{13} c_{23} }  \\
\end{array}} \right) \\ 
 \end{array}
},
\nonumber\\
&&{K'} = diag\left( {e^{i\left( {\beta _1 - \rho } \right) } ,e^{i\beta _2 } ,e^{i
\beta _3 } } \right),
\label{U_PMNS^PDG}
\end{eqnarray}
%%%%%%%%%%%%%%%%%%%%
with $\delta_{CP} = \delta + \rho$ as a Dirac CP violating phase.

We report main results obtained in Ref.\cite{Leptonic CP violation induced}. 
In the next section, we introduce the approximately $\mu-\tau$ symmetry, which 
is related to the invariance
under the interchange of $\nu _{\mu}   \leftrightarrow - \sigma \nu _{\tau} \left( \sigma=\pm1 \right)$.
In paticular, we discuss general phase structure for $N$ subject to the $\mu-\tau$
 symmetry.
In the third section, we argue how we describe the mass hierarchy in our model.
In the fourth section, we estimate the CP violating phases in each mass hierarchy.
The final section is devoted to sammary and discussions.
%%%%%%%%%%%%%%%%%%%%%%%%%%%%%%%%%%%%%%%%%%%%%%%%%%%%%%%%%%%%%%%%%%%%%%%%%%%%%%%%
%%%%%%%%%%%%%%%%%%%%%%%%%%%%%%%%%%%%%%%%%%%%%%%%%%%%%%%%%%%%%%%%%%%%%%%%%%%%%%%%
\section{$\mu-\tau$ symmetry and it's breaking}

The mass matrix is devided into the $\mu-\tau$ symmetric part $M_{\nu}^{\left(  +  \right)}$
 and the $\mu-\tau$ symmetry breaking part $M_{\nu}^{\left(  -  \right)}$:
%%%%%%%%%%%%%%%%%%%%
\begin{eqnarray}
&&M_\nu  = M_\nu  ^{\left(  +  \right)}  + M_\nu  ^{\left(  -  \right)}, 
\nonumber\\
&&
{\scriptsize
M_\nu  ^{\left(  +  \right)}  \equiv \left( {\begin{array}{*{20}c}
   {M_{ee} } & {M_{e\mu } ^{\left(  +  \right)} } & { - \sigma M_{e\mu } ^{\left
(  +  \right)} }  \\
   {M_{e\mu } ^{\left(  +  \right)} } & {M_{\mu \mu } ^{\left(  +  \right)} } & 
{M_{\mu \tau } }  \\
   { - \sigma M_{e\mu } ^{\left(  +  \right)} } & {M_{\mu \tau } } & {M_{\tau 
\tau } ^{\left(  +  \right)} }  \\
\end{array}} \right)
},
\nonumber\\
&&
{\scriptsize
M_\nu  ^{\left(  -  \right)}  \equiv \left( {\begin
{array}{*{20}c}
   0 & {M_{e\mu } ^{\left(  -  \right)} } & {\sigma M_{e\mu } ^{\left(  -  \right)} }  \\
   {M_{e\mu } ^{\left(  -  \right)} } & {M_{\mu \mu } ^{\left(  -  \right)} } & 0  \\
   {\sigma M_{e\mu } ^{\left(  -  \right)} } & 0 & { - M_{\tau \tau } ^{\left(  
-  \right)} }  \\
\end{array}} \right)},
\label{definition of M^+-}
\end{eqnarray}
%%%%%%%%%%%%%%%%%%%%
which is just an identity.
The interchange of $\nu _\mu   \leftrightarrow  - \sigma \nu _\tau$ leads to 
$M_{\nu}^{\left(  \pm  \right)}  \rightarrow  \pm M_{\nu}^{\left(  \pm  \right)}$
 .
We also use the notation:
$a_0=M_{ee},b_0=M_{e\mu}^{\left(+\right)},d_0=M_{\mu\mu}^{\left(+\right)}$ and 
$e_0=M_{\mu\tau}$.
The $\mu-\tau$ symmetric part $M_\nu  ^{\left(  +  \right)}$ gives the mixing angles as:
%%%%%%%%%%%%%%%%%%%%
\begin{equation}
s_{13}  = 0,\,\,\,\,s_{23}  = \frac{\sigma }{{\sqrt 2 }},\,\,\,\,
\tan 2\theta _{12}  = \frac{{2\sqrt 2 b_0 }}{{a_0  - d_0  + \sigma e_0 }}.
\label{good points of mu-tau sym}
\end{equation}
%%%%%%%%%%%%%%%%%%%
However, the Dirac CP phase vanishes because it appears as the coefficient 
of $s_{13}$.

We introduce 
%%%%%%%%%%%%%%%%%%%%
\begin{equation}
\nu _ \pm   = \frac{{\nu _\mu   \pm \left( { - \sigma \nu _\tau  } \right)}}{{\sqrt 2 }},
\,\,\,\,N_ \pm   = \frac{{N_{f_1}   \pm \left( { - \sigma N_{f_2}  } \right)}}{{
\sqrt 2 }},
 \label{nu_pm,N_pm}
\end{equation}
%%%%%%%%%%%%%%%%%%%%
to describe the effect of the $\mu-\tau$ symmetry and it's breaking, where $f_1=\mu$
 and $f_2=\tau$ for $N$ subject to the $\mu-\tau$ symmetry.
$N_{\pm}$, $M_R$ and $Y_{\nu}$ are parametrized as 
%%%%%%%%%%%%%%%%%%%%
{\scriptsize
\begin{eqnarray}
&&Y_\nu   = Y_\nu  ^{\left(  +  \right)}  + Y_\nu  ^{\left(  -  \right)},
\nonumber\\
&&
Y_\nu  ^{\left(  +  \right)}  \equiv \left( {\begin{array}{*{20}c}
   {h_{ + e} ^{\left(  +  \right)} } & {h_{ + \mu } ^{\left(  +  \right)} } & {h_
{ + \tau } ^{\left(  +  \right)} }  \\
   {h_{ - e} ^{\left(  +  \right)} } & {h_{ - \mu } ^{\left(  +  \right)} } & {h_
{ - \tau } ^{\left(  +  \right)} }  \\
\end{array}} \right),
\nonumber\\
&&
Y_\nu  ^{\left(  -  \right)}  \equiv \left( {\begin
{array}{*{20}c}
   {h_{ + e} ^{\left(  -  \right)} } & {h_{ + \mu } ^{\left(  -  \right)} } & {h_
{ + \tau } ^{\left(  -  \right)} }  \\
   {h_{ - e} ^{\left(  -  \right)} } & {h_{ - \mu } ^{\left(  -  \right)} } & {h_
{ - \tau } ^{\left(  -  \right)} }  \\
\end{array}} \right),
\nonumber\\
&& M_R  = M_R ^{\left(  +  \right)}  + M_R ^{\left(  -  \right)},
\nonumber\\
&&  M_R ^{\left(  +  \right)}  \equiv diag\left( {M_{R +  + } ,M_{R -  - } } \right),
\nonumber\\
&&  M_R ^{\left(  -  \right)}  \equiv \left( {\begin{array}{*{20}c}
   0 & {e^{i\Theta _{ +  - } } M_{R +  - } }  \\
   {e^{i\Theta _{ +  - } } M_{R +  - } } & 0  \\
\end{array}} \right),
 \label{U_R and Devided Y_nu and M_R}
\end{eqnarray}
}
%%%%%%%%%%%%%%%%%%%%
where the $\mu-\tau$ symmetrc parts are $Y_{\nu}^{\left(  +  \right)}$ and 
$M_R ^{\left(  +  \right)}$
and the $\mu-\tau$ symmetry breaking parts are $Y_{\nu}^{\left(  -  \right)}$ and 
$M_R ^{\left(  -  \right)}$. The subscripts $\pm$ represent terms for $N_\pm$.

For $N$ subject to the $\mu-\tau$ symmetry, we require that ${M_{R +  - } } \approx 0$,
 ${\left|h_{ij}  ^{\left(  -  \right)} \right|}  \ll 1$
and that the mixing between $N_+$ and $N_-$ be small. We obtain that, for $\omega$
 associated with the $N_+$-$N_-$ mixing,
%%%%%%%%%%%%%%%%%%%%
{\scriptsize
\begin{eqnarray}
&&
 Y_\nu  ^{\left(  +  \right)}  = \left( {\begin{array}{*{20}c}
   {h_{ + e} ^{\left(  +  \right)} } & {h_{ + \mu } ^{\left(  +  \right)} } & { 
- \sigma h_{ + \mu } ^{\left(  +  \right)} }  \\
   0 & {h_{ - \mu } ^{\left(  +  \right)} } & {\sigma h_{ - \mu } ^{\left(  +  
\right)} }  \\
\end{array}} \right),
\nonumber\\
&&
Y_\nu  ^{\left(  -  \right)}  = \left( {\begin
{array}{*{20}c}
   0 & {h_{ + \mu } ^{\left(  -  \right)} } & {\sigma h_{ + \mu } ^{\left(  -  
\right)} }  \\
   {h_{ - e} ^{\left(  -  \right)} } & {h_{ - \mu } ^{\left(  -  \right)} } & { 
- \sigma h_{ - \mu } ^{\left(  -  \right)} }  \\
\end{array}} \right), 
\nonumber\\
&&
M_{ee}  \approx  - v^2 M_1 ^{ - 1} h_{ + e} ^{\left(  +  \right)2},
\nonumber\\
&&
M_{e\mu } ^{\left(  +  \right)}  \approx  - v^2 M_1 ^{ - 1} h_{ + e} ^{\left(  
+  \right)} h_{ + \mu } ^{\left(  +  \right)},
\nonumber\\
&&
M_{e\mu }^{\left(  -  \right)}  \approx  - v^2 \left( \begin{array}{l}
 \left( {h_{ + \mu }^{\left(  -  \right)}  - sh_{ - \mu }^{\left(  +  \right)} e^{ - i\omega } } \right)h_{ + e}^{\left(  +  \right)} M_1^{ - 1}  \\ 
  + \left( {h_{ - e}^{\left(  -  \right)}  + sh_{ + e}^{\left(  +  \right)} e^{i\omega } } \right)h_{ - \mu }^{\left(  +  \right)} M_2^{ - 1}  \\ 
 \end{array} \right),
\nonumber\\
&&
M_{\mu \mu } ^{\left(  +  \right)}  \approx  - v^2 \left( {M_2 ^{ - 1} h_{ - 
\mu } ^{\left(  +  \right)}  + M_1 ^{ - 1} h_{ + \mu } ^{\left(  +  \right)2} } 
\right), 
\nonumber\\
&&
M_{\mu \mu }^{\left(  -  \right)}  \approx  - 2v^2 \left( \begin{array}{l}
 M_2^{ - 1} h_{ - \mu }^{\left(  +  \right)} \left( {h_{ - \mu }^{\left(  -  \right)}  + se^{i\omega } h_{ + \mu }^{\left(  +  \right)} } \right) \\ 
  + M_1^{ - 1} h_{ + \mu }^{\left(  +  \right)} \left( {h_{ + \mu }^{\left(  -  \right)}  - se^{ - i\omega } h_{ - \mu }^{\left(  +  \right)} } \right) \\ 
 \end{array} \right), 
 \nonumber\\
&&
M_{\mu \tau }  \approx  - \left( { - \sigma } \right)v^2 \left( {M_1 ^{ - 1} h_
{ + \mu } ^{\left(  +  \right)2}  - M_2 ^{ - 1} h_{ - \mu } ^{\left(  +  \right)
2} } \right). 
 \label{Masses that N subject to the mu-tau} 
\end{eqnarray}
}
%%%%%%%%%%%%%%%%%%%%
up to the first order of the breaking terms, where $v = \left\langle 0 \right|H_u \left| 0 \right\rangle $
 and $\omega \approx0$.
In this case, we can absorb all phases of $Y_\nu^{ \left( + \right)}$ into phases 
of $\nu_{e,+,-}$, and therefore,
the CP phases only arise from $M_{\nu}^{(-)}$.
We use $M_{\nu}$ by $\alpha=\arg \left( {M_{e\mu } ^{\left(  -  \right)} } \right)$ and 
$\beta=\arg \left( {M_{\mu\mu } ^{\left(  -  \right)} } \right)$.
If we consider the second order of the breaking terms, which have been safely 
neglected, the $\mu-\tau$ symmetric part $M_{\nu}^{\left( + \right)}$ includes 
CP phases.
For $N$ blind to the $\mu-\tau$ symmetry, we obtain that
%%%%%%%%%%%%%%%%%%%%
{\scriptsize
\begin{equation}
\begin{array}{l}
 Y_\nu  ^{\left(  +  \right)}  \equiv \left( {\begin{array}{*{20}c}
   {h_{ + e} ^{\left(  +  \right)} } & {h_{ + \mu } ^{\left(  +  \right)} } & { 
- \sigma h_{ + \mu } ^{\left(  +  \right)} }  \\
   {h_{ - e} ^{\left(  +  \right)} } & {h_{ - \mu } ^{\left(  +  \right)} } & { 
- \sigma h_{ - \mu } ^{\left(  +  \right)} }  \\
\end{array}} \right),\,\,\,\,\,Y_\nu  ^{\left(  -  \right)}  \equiv \left( {\begin
{array}{*{20}c}
   0 & {h_{ + \mu } ^{\left(  -  \right)} } & {\sigma h_{ + \mu } ^{\left(  -  
\right)} }  \\
   0 & {h_{ - \mu } ^{\left(  -  \right)} } & {\sigma h_{ - \mu } ^{\left(  -  
\right)} }  \\
\end{array}} \right), \\ 
M_{ee}  =  - v^2 \left( \begin{array}{l}
 M_1^{ - 1} \left( {h_{ + e}^{\left(  +  \right)} c - h_{ - e}^{\left(  +  \right)} se^{ - i\omega } } \right)^2  \\ 
  + M_2^{ - 1} \left( {h_{ + e}^{\left(  +  \right)} se^{i\omega }  + h_{ - e}^{\left(  +  \right)} c} \right)^2  \\ 
 \end{array} \right)
, \\ 
 M_{e\mu } ^{\left(  +  \right)}  =  - v^2 \left( \begin{array}{l}
 M_1 ^{ - 1} \left( {h_{ + e} ^{\left(  +  \right)} c - h_{ - e} ^{\left(  +  
\right)} se^{ - i\omega } } \right)\left( {ch_{ + \mu } ^{\left(  +  \right)}  - 
se^{ - i\omega } h_{ - \mu } ^{\left(  +  \right)} } \right) \\ 
  + M_2 ^{ - 1} \left( {h_{ + e} ^{\left(  +  \right)} se^{i\omega }  + h_{ - e}
 ^{\left(  +  \right)} c} \right)\left( {se^{i\omega } h_{ + \mu } ^{\left(  +  
\right)}  + ch_{ - \mu } ^{\left(  +  \right)} } \right) \\ 
 \end{array} \right), \\ 
 M_{e\mu } ^{\left(  -  \right)}  =  - v^2 \left( \begin{array}{l}
 M_1 ^{ - 1} \left( {h_{ + e} ^{\left(  +  \right)} c - h_{ - e} ^{\left(  +  
\right)} se^{ - i\omega } } \right)\left( {ch_{ + \mu } ^{\left(  -  \right)}  - 
se^{ - i\omega } h_{ - \mu } ^{\left(  -  \right)} } \right) \\ 
  + M_2 ^{ - 1} \left( {h_{ + e} ^{\left(  +  \right)} se^{i\omega }  + h_{ - e}
 ^{\left(  +  \right)} c} \right)\left( {se^{i\omega } h_{ + \mu } ^{\left(  -  
\right)}  + ch_{ - \mu } ^{\left(  -  \right)} } \right) \\ 
 \end{array} \right), \\ 
M_{\mu \mu }^{\left(  +  \right)}  =  - v^2 \left( \begin{array}{l}
 M_1^{ - 1} \left( {h_{ + \mu }^{\left(  +  \right)} c - se^{ - i\omega } h_{ - \mu }^{\left(  +  \right)} } \right)^2  \\ 
  + M_2^{ - 1} \left( {se^{i\omega } h_{ + \mu }^{\left(  +  \right)}  + ch_{ - \mu }^{\left(  +  \right)} } \right)^2  \\ 
 \end{array} \right)
, \\ 
 M_{\mu \mu } ^{\left(  -  \right)}  =  - 2v^2 \left( \begin{array}{l}
 M_1 ^{ - 1} \left( {ch_{ + \mu } ^{\left(  +  \right)}  - se^{ - i\omega } h_{ 
- \mu } ^{\left(  +  \right)} } \right)\left( {ch_{ + \mu } ^{\left(  -  \right)
}  - se^{ - i\omega } h_{ - \mu } ^{\left(  -  \right)} } \right) \\ 
  + M_2 ^{ - 1} \left( {se^{i\omega } h_{ + \mu } ^{\left(  +  \right)}  + ch_{ 
- \mu } ^{\left(  +  \right)} } \right)\left( {se^{i\omega } h_{ + \mu } ^{\left
(  -  \right)}  + ch_{ - \mu } ^{\left(  -  \right)} } \right) \\ 
 \end{array} \right), \\ 
M_{\mu \tau }  =  - \left( { - \sigma } \right)v^2 \left( \begin{array}{l}
 M_1^{ - 1} \left( {ch_{ + \mu }^{\left(  +  \right)}  - se^{ - i\omega } h_{ - \mu }^{\left(  +  \right)} } \right)^2  \\ 
  + M_2^{ - 1} \left( {ch_{ - \mu }^{\left(  +  \right)}  + se^{i\omega } h_{ + \mu }^{\left(  +  \right)} } \right)^2  \\ 
 \end{array} \right), 
 \end{array}
 \label{Masses that N blind to the mu-tau} 
\end{equation}
 }
%%%%%%%%%%%%%%%%%%%%
where $\theta$ is the $N_+$-$N_-$ mixing angle ($c=\cos\theta$, $s=\sin\theta$),
up to the first order of the breaking terms of 
$h_{ - e} ^{\left(  -  \right)} ,h_{ + \mu } ^{\left(  -  \right)}$ and $h_{ - \mu } ^{\left(  -  \right)} $.
The $\mu-\tau$ symmetric part of $M_{\nu}^{\left( + \right)}$ is not generally real.
To performed similary consideration, we assume the same phase structure as the 
one for $N$ subject to the $\mu-\tau$ symmetry, namely , $M_\nu^{\left(  + \right)}$
 is real.

We obtain neutrino masses from Eq.(\ref{m_1 m_2 m_3}) as
%%%%%%%%%%%%%%%%%%%%
\begin{equation}
\begin{array}{l}
\begin{array}{l}
 m_1 e^{ - 2i\beta _1 }  \approx \frac{1}{2}\left[ \begin{array}{l}
 e^{2i\rho } a_0  + d_0  - \sigma e_0  \\ 
  + 2\left( {2i\Delta \gamma d_0  + \left( {i\gamma  + \Delta } \right)d_{0'} e^{i\beta } } \right) \\ 
 \end{array} \right] \\ 
 {\rm{     }}\,\, - \frac{{\sqrt 2 }}{{\sin 2\theta _{12} }}\left\{ {b_0 \left( {1 + i\Delta \gamma } \right) + b_{0'} e^{i\alpha } \left( {\Delta  + i\gamma } \right)} \right\} \\ 
 \end{array}, \\ 
\begin{array}{l}
 m_2 e^{ - 2i\beta _2 }  \approx \frac{1}{2}\left[ \begin{array}{l}
 e^{2i\rho } a_0  + d_0  - \sigma e_0  \\ 
  + 2\left( {2i\Delta \gamma d_0  + \left( {i\gamma  + \Delta } \right)d_{0'} e^{i\beta } } \right) \\ 
 \end{array} \right] \\ 
 {\rm{     }} + \frac{{\sqrt 2 }}{{\sin 2\theta _{12} }}\left\{ {b_0 \left( {1 + i\Delta \gamma } \right) + b_{0'} e^{i\alpha } \left( {\Delta  + i\gamma } \right)} \right\} \\ 
 \end{array}, \\ 
m_3 e^{ - 2i\beta _3 }  \approx \left[ \begin{array}{l}
 d_0  + \sigma e \\ 
  - 2\left( {i\gamma \Delta 2d_0  - d'e^{i\beta } \left( {i\gamma  - \Delta } \right)} \right) \\ 
 \end{array} \right], 
 \end{array}
 \label{approximated m_1 m_2 m_3} 
\end{equation}
%%%%%%%%%%%%%%%%%%%%
where we have parameterized $M_\nu  ^{\left(  -  \right)}$ to be:
%%%%%%%%%%%%%%%%%%%%
\begin{equation}
M_\nu  ^{\left(  -  \right)}  = \left( {\begin{array}{*{20}c}
   0 & {b_0'e^{i\alpha } } & {\sigma b_0'e^{i\alpha } }  \\
   {b_0'e^{i\alpha } } & {d_0'e^{i\beta } } & 0  \\
   {\sigma b_0'e^{i\alpha } } & 0 & { - d_0'e^{i\beta } }  \\
\end{array}} \right).
\label{parameterized Mnu^-}
\end{equation}
%%%%%%%%%%%%%%%%%%%%
The parameter $\Delta$ is defined by $s_{23}  = \sigma {{\left( {1 - \Delta } \right)} \mathord{\left/ {\vphantom {{\left( {1 - \Delta } \right)} {\sqrt 2 }}} \right. \kern-\nulldelimiterspace} {\sqrt 2 }},\,\,\,\,c_{23}  = {{\left( {1 + \Delta } \right)} \mathord{\left/ {\vphantom {{\left( {1 + \Delta } \right)} {\sqrt 2 }}} \right. \kern-\nulldelimiterspace} {\sqrt 2 }}$.
For the normal mass hierarchy, $m_2$ can be further converted into
%%%%%%%%%%%%%%%%%%%%
\begin{equation}
m_2 e^{ - 2i\beta _2 }  \approx \frac{{2\sqrt 2 }}{{\sin 2\theta _{12} }}\left\{ 
{b_0 \left( {1 + i\Delta \gamma } \right) + b_0 'e^{i\alpha } \left( {\Delta  + 
i\gamma } \right)} \right\},
 \label{approximated m_2 of normal} 
\end{equation}
%%%%%%%%%%%%%%%%%%%%
 because of $m_1 = 0$.
%%%%%%%%%%%%%%%%%%%%%%%%%%%%%%%%%%%%%%%%%%%%%%%%%%%%%%%%%%%%%%%%%%%%%%%%%%%%%%%%
%%%%%%%%%%%%%%%%%%%%%%%%%%%%%%%%%%%%%%%%%%%%%%%%%%%%%%%%%%%%%%%%%%%%%%%%%%%%%%%%
%%%%%%%%%%%%%%%%%%%%%%%%%%%%%%%%%%%%%%%%%%%%%%%%%%%%%%%%%%%%%%%%%%%%%%%%%%%%%%%%
\section{Describing the mass hierarchy}
In this section, we argue how the mass hierarchy is described
by using Eq.(\ref{Masses that N subject to the mu-tau}) and Eq.
(\ref{Masses that N blind to the mu-tau}).
To perform numerical calculation, we have assumed $\omega=0$ and the tri-bi maximal mixing for 
$M_\nu^{\left(  + \right)}$ \cite{Tri-bi maximal}
which gives 
$\sin ^2 \theta _{12}  = {1 \mathord{\left/ {\vphantom {1 3}} \right. \kern-\nulldelimiterspace} 3},\,\,\,\sin ^2 \theta _{23}  = {1 \mathord{\left/ {\vphantom {1 2}} \right. \kern-\nulldelimiterspace} 2},\,\,\,\sin ^2 \theta _{13}  = 0$.
There are three textures.
These textures can reproduce respectively the normal mass hierarchy,
 the inverted mass hierarchy with ${m_1  \approx m_2 }$ and the inverted mass hierarchy 
with ${m_1  \approx - m_2 }$.
%%%%%%%%%%%%%%%%%%%%%%%%%%%%%%%%%%%%%%%%
%%%%%%%%%%%%%%%%%%%%%%%%%%%%%%%%%%%%%%%%%%%%%%%%%%%%%%%%%%%% Normal
%%%%%%%%%%%%%%%%%%%%%%%%%%%%%%%%%%%%%%%%
\subsection{Normal mass hierarchy($ m_2  \ll m_3 $)}
As we can see from Eq.(\ref{Masses that N subject to the mu-tau}), we have the 
relation for $N$ subject to $\mu-\tau$ symmetry:
%%%%%%%%%%%%%%%%%%%%
\begin{equation}
M_{ee} M_{\mu \mu } ^{\left(  +  \right)}  = \left( {M_{e\mu } ^{\left(  +  \right)
} } \right)^2
 \label{relation of masses for N subject} 
\end{equation}
%%%%%%%%%%%%%%%%%%%%
up to the first order of $\mu-\tau$ symmetry breakings.
This relation forbids us to use the inverted mass hierarchy because we can not 
reproduce the $\mu-\tau$ symmetric mass matrix 
for the inverted mass hierarchy.  We see that
$M_{e\mu } ^{\left(  +  \right)} \approx  0$ and 
$M_{ee }, M_{\mu\mu } ^{\left(  +  \right)} \not  \approx  0$
 for the inverted mass hierarchy with ${m_1  \approx  m_2 }$ and that
$M_{ee}M_{\mu\mu } ^{\left(  +  \right)} < 0$ for the inverted mass hierarchy 
with ${m_1  \approx - m_2 }$.
We are only allowed to use the normal mass hierarchy.

The normal mass hierarchy is realized \cite{Textures} by 
%%%%%%%%%%%%%%%%%%%%
\begin{equation}
M_\nu  ^{\left(  +  \right)}  = m_0 \left( {\begin{array}{*{20}c}
   {p\eta } & \eta  & { - \eta }  \\
   \eta  & 1 & {1 - s\eta }  \\
   { - \eta } & {1 - s\eta } & 1  \\
\end{array}} \right),
\label{normal texture}
\end{equation}
%%%%%%%%%%%%%%%%%%%%
for $\left| \eta  \right|  \ll 1$.
The condition $\det\left( M_{\nu} \right) =0$ requires the relation 
$ {p = {2 \mathord{\left/ {\vphantom {2 s}} \right. \kern-\nulldelimiterspace} s}} $.
We require that $s=2$ for the tri-bi maximal mixing. The neutrino masses are 
computed from Eq.(\ref{approximated m_1 m_2 m_3}) and Eq.(\ref{approximated m_2 of normal})
 giving
%%%%%%%%%%%%%%%%%%%%
\begin{equation}
\begin{array}{l}
 m_2 e^{ - 2i\beta _2 }  \approx \frac{{2\sqrt 2 \eta e^{i\rho } }}{{\sin 2\theta 
_{12} }}m_0, \\ 
m_3 e^{ - 2i\beta _3 }  \approx \left( \begin{array}{l}
 m_0 \left( {2 - s\eta } \right) \\ 
  - 2\left( {2i\gamma \Delta m_0  + d'_0 \left( {\Delta  - i\gamma } \right)e^{i\beta } } \right) \\ 
 \end{array} \right). 
 \end{array}
 \label{normal mass m_2 and m_3} 
\end{equation}
%%%%%%%%%%%%%%%%%%%%%%%%%%%%%%%%%%%%%%%%
%%%%%%%%%%%%%%%%%%%%%%%%%%%%%%%%%%%%%%%%%%%%%%%%%%%%%%%%%%%% All Inverted 
%%%%%%%%%%%%%%%%%%%%%%%%%%%%%%%%%%%%%%%%
\subsection{Inverted mass hierarchy}
The Eq.(\ref{Masses that N blind to the mu-tau}) gives us the similar relation 
to Eq.(\ref{relation of masses for N subject})
for $N$ blind to the $\mu-\tau$ symmetry:
%%%%%%%%%%%%%%%%%%%%
\begin{equation}
M_{\mu \mu } ^{\left(  +  \right)}  =  - \sigma M_{\mu \tau } ,
 \label{relation of masses for N blind} 
\end{equation}
%%%%%%%%%%%%%%%%%%%%
leading to $m_3 = 0$.
Therefore Eq.(\ref{relation of masses for N blind}) only allows the inverted mass 
hierarchy with either $m_1  \approx \pm m_2 $.
%%%%%%%%%%%%%%%%%%%%%%%%%%%%%%%%%%%%%%%%
%%%%%%%%%%%%%%%%%%%%%%%%%%%%%%%%%%%%%%%% Inverted ‡T
%%%%%%%%%%%%%%%%%%%%%%%%%%%%%%%%%%%%%%%%

\noindent
{\bf{Type-I} (${m_1  \approx m_2 }$)\\}
The type-I is realized \cite{Textures} by
%%%%%%%%%%%%%%%%%%%%
\begin{equation}
M_\nu  ^{\left(  +  \right)}  = m_0 \left( {\begin{array}{*{20}c}
   {2 - p\eta } & \eta  & { - \sigma \eta }  \\
   \eta  & 1 & { - \sigma }  \\
   { - \sigma \eta } & { - \sigma } & 1  \\
\end{array}} \right),
 \label{inverted1 texture} 
\end{equation}
%%%%%%%%%%%%%%%%%%%%
for $p=1$ to describe the tri-bi maximal mixing and $\left| \eta \right|  \ll  1$.
The neutrino masses are computed from Eq.(\ref{approximated m_1 m_2 m_3}) to be:
%%%%%%%%%%%%%%%%%%%%
\begin{equation}
\begin{array}{l}
 m_1 e^{ - 2i\beta _1 }  \approx m_0 \left( {1 + e^{2i\rho }  - \frac{1}{2}p\eta 
e^{2i\rho }  - \frac{{\sqrt 2 e^{i\rho } \eta }}{{\sin 2\theta _{12} }}} \right)
, \\ 
 m_2 e^{ - 2i\beta _2 }  \approx m_0 \left( {1 + e^{2i\rho }  - \frac{1}{2}p\eta 
e^{2i\rho }  + \frac{{\sqrt 2 e^{i\rho } \eta }}{{\sin 2\theta _{12} }}} \right)
. \\ 
 \end{array}
 \label{inverted1 mass m_1 and m_2} 
\end{equation}
%%%%%%%%%%%%%%%%%%%%
%%%%%%%%%%%%%%%%%%%%%%%%%%%%%%%%%%%%%%%%
%%%%%%%%%%%%%%%%%%%%%%%%%%%%%%%%%%%%%%%% Inverted ‡U
%%%%%%%%%%%%%%%%%%%%%%%%%%%%%%%%%%%%%%%%
\noindent
{\bf{Type-I\hspace{-.1em}I} (${m_1  \approx - m_2 }$)}\\

The type-I\hspace{-.1em}I is realized \cite{Textures} by
%%%%%%%%%%%%%%%%%%%%
\begin{equation}
M_\nu  ^{\left(  +  \right)}  = m_0 \left( {\begin{array}{*{20}c}
   { - 2 + \eta } & q & { - \sigma q}  \\
   q & 1 & { - \sigma }  \\
   { - \sigma q} & { - \sigma } & 1  \\
\end{array}} \right),
\label{inverted2 texture} 
\end{equation}
%%%%%%%%%%%%%%%%%%%%
for $q=4$ to describe the tri-bi maximal mixing and $\left| \eta \right|  \ll  1$.
The neutrino masses are also computed from Eq.(\ref{approximated m_1 m_2 m_3}) to be:
%%%%%%%%%%%%%%%%%%%%
\begin{equation}
\begin{array}{l}
  m_1 e^{ - 2i\beta _1 }  \approx m_0 \left\{ {\frac{{2\left( {1 - e^{2i\rho } }
 \right) + \eta e^{2i\rho } }}{2} - \frac{{\sqrt 2 e^{i\rho } q}}{{\sin 2\theta 
_{12} }}} \right\}, \\ 
  m_2 e^{ - 2i\beta _2 }  \approx m_0 \left\{ {\frac{{2\left( {1 - e^{2i\rho } }
 \right) + \eta e^{2i\rho } }}{2} + \frac{{\sqrt 2 e^{i\rho } q}}{{\sin 2\theta 
_{12} }}} \right\}.
 \end{array}
 \label{inverted2 mass m_1 and m_2} 
\end{equation}
%%%%%%%%%%%%%%%%%%%%
%%%%%%%%%%%%%%%%%%%%%%%%%%%%%%%%%%%%%%%%%%%%%%%%%%%%%%%%%%%%%%%%%%%%%%%%%%%%%%%%
%%%%%%%%%%%%%%%%%%%%%%%%%%%%%%%%%%%%%%%%%%%%%%%%%%%%%%%%%%%%%%%%%%%%%%%%%%%%%%%%
%%%%%%%%%%%%%%%%%%%%%%%%%%%%%%%%%%%%%%%%%%%%%%%%%%%%%%%%%%%%%%%%%%%%%%%%%%%%%%%%
%%%%%%%%%%%%%%%%%%%%%%%%%%%%%%%%%%%%%%%%%%%%%%%%%%%%%%%%%%%%%%%%%%%%%%%%%%%%%%%%
\section{Estimation of CP violating phases}
In this section, we estimate sizes of CP phases.
Our seesaw model has four phases: three phases from Yukawa couplings and one phase 
from Majorana phase for heavy neurinos.
These phases correspond to one Dirac phase and three Majorana phases.

The CP violating phases of $\delta$ and $\rho$ depend on $X$ and $Y$ 
\cite{Dependance of CP phase on X and Y} as (\ref{dependance of delta and rho}),
which can be calculated to be:
%%%%%%%%%%%%%%%%%%%%
\begin{equation}
{\scriptsize 
\begin{array}{l}
c_{13} X \approx \sqrt 2 \left[ \begin{array}{l}
 b_0 \left( {a_0  - \sigma e_0  + d_0 } \right) + b_{0'} d_{0'} e^{i\left( {\beta  - \alpha } \right)}  \\ 
  + \left( {\Delta  + i\gamma } \right)\left( \begin{array}{l}
 \left( {\sigma e_0  + d_0 } \right)b_{0'} e^{ - i\alpha }  \\ 
  + a_0 b_{0'} e^{i\alpha }  + b_0 d_{0'} e^{i\beta }  \\ 
 \end{array} \right) \\ 
 \end{array} \right], 
\\ 
Y \approx \sqrt 2 \sigma \left( \begin{array}{l}
 \left( {i\gamma  - \Delta } \right)\left( {b_0 \left( {a_0  - \sigma e_0  + d_0 } \right) + b_{0'} d_{0'} e^{i\left( {\beta  - \alpha } \right)} } \right) \\ 
  + \left( {\left( {\sigma e_0  + d_0 } \right)b_{0'} e^{ - i\alpha }  + a_0 b_{0'} e^{i\alpha }  + b_0 d_{0'} e^{i\beta } } \right) \\ 
 \end{array} \right), \\ 
 \end{array}
 }
\label{The approximated X and Y} 
\end{equation}
%%%%%%%%%%%%%%%%%%%%
where we neglect the second order of $\gamma$ and $\Delta$ because of 
$\left|\gamma\right|,\left|\Delta\right| \ll 1$.
From Eq.(\ref{dependance of delta and rho}), we expect that $\rho$ will be suppressed 
if the term of $b_0 \left( {a_0  - \sigma e_0  + d_0 } \right)$ in X is not suppressed
and this is the case of the present discussions.

We will show that the Dirac CP violating phase $\delta_{CP}$ depends on $\delta$ and that
$\rho$ is suppressed because $X$ is almost real. Please see 
the detailed numerical results in \cite{Leptonic CP violation induced}.
%%%%%%%%%%%%%%%%%%%%%%%%%%%%%%%%%%%%%%%%
%%%%%%%%%%%%%%%%%%%%%%%%%%%%%%%%%%%%%%%%%%%%%%%%%%%%%%%%%%%% Normal
%%%%%%%%%%%%%%%%%%%%%%%%%%%%%%%%%%%%%%%%
\subsection{Normal mass hierarchy}
We obtain from (\ref{normal texture}): 
%%%%%%%%%%%%%%%%%%%%
\begin{equation}
\begin{array}{l}
c_{13} X \approx m_0^2 \sqrt 2 \left[ \begin{array}{l}
 \eta ^2 \left( {p + s} \right) + b'_0 b'_0 e^{i\left( {\beta  - \alpha } \right)}  \\ 
  + 2\left( {\Delta  + i\gamma } \right)b'_0 e^{ - i\alpha }  \\ 
 \end{array} \right], \\ 
 Y \approx \sigma 2\sqrt 2 m_0 ^2 b'_0 e^{ - i\alpha },  \\ 
 \end{array}
\label{X and Y in normal}
\end{equation}
%%%%%%%%%%%%%%%%%%%%
from which we can understand the main sources of CP phases as
%%%%%%%%%%%%%%%%%%%%
\begin{equation}
\begin{array}{l}
\rho  \approx \arg \left( \begin{array}{l}
 m_0 ^2 \eta ^2 \left( {p + s} \right) + b_0 'd_0 'e^{i\left( {\beta  - \alpha }
 \right)}  \\ 
  + 2m_0 \left( {\Delta  + i\gamma } \right)b_0 'e^{ - i\alpha }  \\ 
 \end{array} \right) \approx 0,\,\,\,\,\, \delta  \approx \alpha.
 \end{array}
\label{dependance of delta and rho in normal}
\end{equation}\\
%%%%%%%%%%%%%%%%%%%%
However, it is numerically underdtood that $\rho$ will be suppress because of 
$\left|\eta\right| \approx O\left( {0.1} \right) \gg \left| b_0 '\right| \approx \left| d_0 '\right|$.
The Majorana phase is defined by $\varphi  = \frac{1}{2}\left( {\beta _2  - \beta _3 } \right)$
 because $\beta _1$ is irrelevant owing to $m_1=0$. 
From (\ref{normal mass m_2 and m_3}), $\varphi$ is calculated to be:
%%%%%%%%%%%%%%%%%%%%
\begin{equation}
\varphi \approx  - \frac{1}{4}\rho.
\label{dependance of majorana phase in normal}
\end{equation}
%%%%%%%%%%%%%%%%%%%%

%%%%%%%%%%%%%%%%%%%%%%%%%%%%%%%%%%%%%%%%
%%%%%%%%%%%%%%%%%%%%%%%%%%%%%%%%%%%%%%%%%%%%%%%%%%%%%%%%%%%% All Inverted 
%%%%%%%%%%%%%%%%%%%%%%%%%%%%%%%%%%%%%%%%
\subsection{Inverted mass hierarchy}
In this case, we denote the majorana phase by $\varphi  = \frac{1}{2}\left( {\beta _1  - \beta _2 } \right)$.

%%%%%%%%%%%%%%%%%%%%%%%%%%%%%%%%%%%%%%%%
%%%%%%%%%%%%%%%%%%%%%%%%%%%%%%%%%%%%%%%% Inverted‡T
%%%%%%%%%%%%%%%%%%%%%%%%%%%%%%%%%%%%%%%%
\noindent
{\bf{Type-I}\\}
We approximately obtain $X$ and $Y$ from (\ref{inverted1 texture}) as
%%%%%%%%%%%%%%%%%%%%
\begin{equation}
 c_{13} X \approx m_0 ^2 4\sqrt 2 \eta, \,\,\,\,\,\, 
 Y \approx \sigma m_0 ^2 2\sqrt 2 b_0 'e^{i\alpha },  \\ 
\label{X and Y in inverted1}
\end{equation}
%%%%%%%%%%%%%%%%%%%%
leading to
%%%%%%%%%%%%%%%%%%%%
\begin{equation}
  \delta  \approx \alpha,\,\,\,\,\,
  \rho  \approx 0,
\label{dependance of delta and rho in inverted1}
\end{equation}
%%%%%%%%%%%%%%%%%%%%
which result in
%%%%%%%%%%%%%%%%%%%%
\begin{equation}
\varphi  \approx  0 .
\label{dependance of majorana phase in inverted1}
\end{equation}
%%%%%%%%%%%%%%%%%%%%
%%%%%%%%%%%%%%%%%%%%%%%%%%%%%%%%%%%%%%%%
%%%%%%%%%%%%%%%%%%%%%%%%%%%%%%%%%%%%%%%% Inverted ‡U
%%%%%%%%%%%%%%%%%%%%%%%%%%%%%%%%%%%%%%%%

\noindent
{\bf{Type-I\hspace{-.1em}I}\\}
We approximately obtain $X$ and $Y$ from (\ref{inverted2 texture}) as
%%%%%%%%%%%%%%%%%%%%
\begin{equation}
\begin{array}{l}
  c_{13} X \approx \sqrt 2 m_0 ^2 q\eta,  \\
  Y \approx \sigma \sqrt 2 m_0 ^2 \left( {qd_0 'e^{i\beta }  - 2b_0 'e^{i\alpha }
 } \right),
 \end{array} 
\label{X and Y in inverted2}
\end{equation}
%%%%%%%%%%%%%%%%%%%%
leading to
%%%%%%%%%%%%%%%%%%%%
\begin{equation}
\delta  \approx \arg \left( { \sigma \sqrt 2 m_0 ^2 \left( {qd_0 'e^{i\beta }  - 
2b_0 'e^{i\alpha } } \right)} \right),\,\,\,\,\,
\rho  \approx 0.
\label{dependance of delta and rho in inverted2}
\end{equation}
%%%%%%%%%%%%%%%%%%%%
where $\rho$ is also suppressed in this case .
From Eq.(\ref{inverted2 mass m_1 and m_2}), we obtain the majorana phase as
%%%%%%%%%%%%%%%%%%%%
\begin{equation}
\varphi  \approx  - \frac{1}{6}\rho .
\label{dependance of majorana phase in inverted2}
\end{equation}
In all three cases, the Majorana CP violating phase is found to be suppreesed.

%%%%%%%%%%%%%%%%%%%%
%%--------------------------------
%%
%%--------------------------------
%%%%%%%%%%%%%%%%%%%%%%%%%%%%%%%%%%%%%%%%%%%%%%%%%%%%%%%%%%%%%%%%%%%%%%%%%%%%%%%%\section{summary and discussions}
We have discussed how the size of leptonic CP violation is estimated from phases 
of flavor neutrino masses in the approximately $\mu-\tau$ symmetric
minimal seesaw model.
We have also discussed that, if N is subject to the $\mu-\tau$ symmetry, neutrinos 
show the normal mass hierarchy while, if $N$ is blind to the $\mu-\tau$
symmetry, neutrinos show the inverted mass hierarchy.
The clear dependance of leptonic CP violation on phases of flavor neutrino masses 
is presented in Eq.(\ref{The approximated X and Y}) for Dirac CP violation described
by $\delta_{CP}=\delta+\rho$ with $\delta=-\arg\left( Y \right)$ and 
$\rho=-\arg\left( X \right)$ and in Eqs.(\ref{approximated m_1 m_2 m_3})
and (\ref{approximated m_2 of normal}) for Majorana CP violation described by $\varphi$ .
Since $X$ is found to be almost real (while $Y$ becomes generically complex), we 
observe that $\rho$ is suppressed.
We have obtain that  $\varphi \approx  - \rho/4$
 for the normal mass hierarchy,
$\varphi \approx 0$ up to $\rho^2$ for the inverted mass hierarchy with 
$m_1 \approx m_2$ and
$\varphi \approx  - \rho/6$
 for the inverted mass hierarchy with $m_1 \approx - m_2$
as long as the $\mu-\tau$ symmetric flavor neutrino masses are taken to be real.
It should be noted that the real $\mu-\tau$ symmetric flavor neutrino masses arise 
as a general property of neutrinos exhibiting the normal mass hierarchy.

Since Majorana phase turns out to be suppressed because of $\rho \approx 0$, to 
expect larger Majorana CP violation, we may discuss another realization
of the $\mu-\tau$  symmetry for neutrinos.  For the same 
$M_{\nu}$ Eq.(\ref{M_nu}), 
$\sin\theta_{23} \approx  - \sigma/\sqrt 2$
 is taken
instead of the present value of 
$\sin\theta_{23} \approx   \sigma/\sqrt 2$
 \cite{Textures}.
In this case, the larger Majorana CP violation is expected because the role of 
$X$ and $Y$ is almost interchanged and we can find that $X$ becomes complex,
yielding larger $\rho$, thereby, giving larger $\varphi$, and $Y$ becomes almost real.
 The detailed discussions based on this expectation will be presented elsewhere 
\cite{Future work2}.
\begin{acknowledgments}
The author would like to thank Professor M. Yasu\`{e} for  his patient reading of the manuscript and useful advices. 
\end{acknowledgments}
%%%%%%%%%%%%%%%%%%%%%%%%%%%%%%%
%%%%%%%%%%%%%%%%%%%%%%%%%%%%%%%
%%%%%%%%%%%%%%%%%%%%%%%%%%%%%%%
\begin{appendix}
\appendix
\section{\label{sec:Appendix1}: Useful formula}
There are useful formula \cite{Dependance of CP phase on X and Y} for
%%%%%%%%%%%%%%%%%%%%
\begin{equation}
\textbf{ M } \equiv M_{\nu}^{\dagger}M_{\nu}=
\left( {\begin{array}{*{20}c}
   A & B & C  \\
   {B^ *  } & D & E  \\
   {C^ *  } & {E^ *  } & F  \\
\end{array}} \right),
\label{M^daggerM}
\end{equation}
%%%%%%%%%%%%%%%%%%%%
which is expressed to be:
\begin{equation}
{\scriptsize 
\begin{array}{l}
  \textbf{ M } = \textbf{ M }^{\left(  +  \right)} {\rm{ + }}\,\textbf{ M }^{\left
(  -  \right)},  \\
  \textbf{ M }^{\left(  +  \right)}  = \left( {\begin{array}{*{20}c}
   A & {B_ +  } & { - \sigma B_ +  }  \\
   {B_ +  ^ *  } & {D_ +  } & {E_ +  }  \\
   { - \sigma B_ +  ^ *  } & {E_ +  } & {D_ +  }  \\
\end{array}} \right),
\\
\textbf{ M }^{\left(  -  \right)}  = \left( {\begin
{array}{*{20}c}
   0 & {B_ -  } & {\sigma B_ -  }  \\
   {B_ -  ^ *  } & {D_ -  } & {iE_ -  }  \\
   {\sigma B_ -  ^ *  } & { - iE_ -  } & { - D_ -  }  \\
\end{array}} \right), \\
  A = \left| {M_{ee} } \right|^2  + 2\left( {\left| {M_{e\mu } ^{\left(  +  \right)
} } \right|^2  + \left| {M_{e\mu } ^{\left(  -  \right)} } \right|^2 } \right),
\\
B_ +   = M_{ee} ^* M_{e\mu } ^{\left(  +  \right)}  + M_{e\mu } ^{\left(  
+  \right)*} \left( {M_{\mu \mu } ^{\left(  +  \right)}  - \sigma M_{\mu \tau } }
 \right) + M_{e\mu } ^{\left(  -  \right)*} M_{\mu \mu } ^{\left(  -  \right)}, 
\\
  D_ +   = \left| {M_{e\mu } ^{\left(  +  \right)} } \right|^2  + \left| {M_{e
\mu } ^{\left(  -  \right)} } \right|^2  + \left| {M_{\mu \mu } ^{\left(  +  
\right)} } \right|^2  + \left| {M_{\mu \mu } ^{\left(  -  \right)} } \right|^2  
+ \left| {M_{\mu \tau } } \right|^2,  \\
  E_ +   = {\mathop{\rm Re}\nolimits} \left( E \right) = \sigma \left( {\left| 
{M_{e\mu } ^{\left(  -  \right)} } \right|^2  - \left| {M_{e\mu } ^{\left(  +  
\right)} } \right|^2 } \right) + 2{\mathop{\rm Re}\nolimits} \left( {M_{\mu \mu }
 ^{\left(  +  \right)*} M_{\mu \tau } } \right), \\
  B_ -   = M_{ee} ^* M_{e\mu } ^{\left(  +  \right)}  + M_{e\mu } ^{\left(  -  
\right)*} \left( {M_{\mu \mu } ^{\left(  +  \right)}  - \sigma M_{\mu \tau } } 
\right) + M_{e\mu } ^{\left(  +  \right)*} M_{\mu \mu } ^{\left(  -  \right)},  
\\
  D_ -   = 2{\mathop{\rm Re}\nolimits} \left( {M_{e\mu } ^{\left(  -  \right)*} 
M_{e\mu } ^{\left(  +  \right)}  + M_{\mu \mu } ^{\left(  -  \right)*} M_{\mu 
\mu } ^{\left(  +  \right)} } \right),
\\
  E_ -   = {\mathop{\rm Im}\nolimits} \left( E \right) = 2{\mathop{\rm Im}\nolimits}
 \left( {M_{e\tau } M_{\mu \mu } ^{\left(  -  \right)*}  - \sigma M_{e\mu } ^{
\left(  -  \right)*} M_{e\mu } ^{\left(  +  \right)} } \right), \\
 \end{array}
}
\label{M^daggerM ^+ ,^- and components}
\end{equation}
%%%%%%%%%%%%%%%%%%%%
$\left(  +  \right)$ and $\left(  -  \right)$, respectively, where indicated the $\mu-\tau$
 symmetry conserving terms and breaking terms.
We then obtain formula for three neutrino angles and Dirac CP violating phases,
%%%%%%%%%%%%%%%%%%%%
\begin{equation}
\begin{array}{l}
 \tan 2\theta _{12} e^{i\rho }  = \frac{2}{{\Lambda _2  - \Lambda _1 }}X,\,\,\,\,
\,\tan 2\theta _{13} e^{ - i\delta }  = \frac{2}{{\Lambda _3  - A}}Y, \\ 
 {\mathop{\rm Re}\nolimits} \left( {E'} \right)\cos 2\theta _{23}  + D_ -  \sin 
2\theta _{23}  + i{\mathop{\rm Im}\nolimits} \left( {E'} \right) =  - s_{13} e^{i
\left( {\rho  + \delta } \right)} X, \\ 
 X = \frac{{c_{23} B - s_{23} C}}{{c_{13} }},\,\,\,\,\,Y = s_{23} B + c_{23} C, \\
 \Lambda _1  \equiv \frac{{c_{13} ^2 A - s_{13} ^2 \Lambda _3 }}{{c_{13} ^2  - s_
{13} ^2 }} \\ 
 \Lambda _2  \equiv c_{23} ^2 D + s_{23} ^2 F - 2s_{23} c_{23} {\mathop{\rm Re}
\nolimits} \left( {E'} \right) \\ 
 \Lambda _3  \equiv s_{23} ^2 D + c_{23} ^2 F + 2s_{23} c_{23} {\mathop{\rm Re}
\nolimits} \left( {E'} \right) \\ 
\end{array}
\label{three useful formula}
\end{equation}
%%%%%%%%%%%%%%%%%%%%
where
%%%%%%%%%%%%%%%%%%%%
\begin{equation}
\arg \left( X \right) = \rho ,\,\,\,\,\arg \left( Y \right) =  - \delta. 
\label{dependance of delta and rho}
\end{equation}
%%%%%%%%%%%%%%%%%%%%
The neutrino masses are computed as 
%%%%%%%%%%%%%%%%%%%%
\begin{equation}
\begin{array}{l}
 m_1 e^{ - 2i\beta _1 }  = \frac{1}{2}\left( {\lambda _1 ' + \lambda _2 e^{ - 2i
\rho } } \right) - \frac{{xe^{ - i\rho } }}{{\sin 2\theta _{12} }}, \\ 
 m_2 e^{ - 2i\beta _2 }  = \frac{1}{2}\left( {\lambda _1 'e^{2i\rho }  + \lambda 
_2 } \right) + \frac{{xe^{i\rho } }}{{\sin 2\theta _{12} }}, \\ 
 m_3 e^{ - 2i\beta _3 }  = \frac{1}{2}\left( {ae^{ - 2i\delta }  + \lambda _3 } 
\right) + \frac{{\lambda _3  - ae^{ - 2i\delta } }}{{2\cos 2\theta _{13} }}, \\ 
 \end{array}
\label{m_1 m_2 m_3}
\end{equation}
%%%%%%%%%%%%%%%%%%%%
where
%%%%%%%%%%%%%%%%%%%%
\begin{equation}
\begin{array}{l}
  \lambda _1  = e^{2i\rho } \frac{{c_{13} ^2 a - s_{13} ^2 e^{2i\delta } \lambda 
_3 }}{{c_{13} ^2  - s_{13} ^2 }}, \\
  \lambda _2  = c_{23} ^2 e^{2i\gamma } d + s_{23} ^2 e^{ - 2i\gamma } f - 2s_{23}
 c_{23} e, \\
  \lambda _3  = s_{23} ^2 e^{2i\gamma } d + c_{23} ^2 e^{ - 2i\gamma } f + 2s_{23}
 c_{23} e, \\ 
  x = \frac{1}{{c_{13} }}e^{i\rho } \left( {c_{23} e^{i\gamma } b - s_{23} e^{ - 
i\gamma } c} \right).
 \end{array}
\label{lambda_1,2,3 and x}
\end{equation}
\end{appendix}
%%%%%%%%%%%%%%%%%%%%%%%%%%%%%%%
%%%%%%%%%%%%%%%%%%%%%%%%%%%%%%%
%%%%%%%%%%%%%%%%%%%%%%%%%%%%%%%
\bigskip % extra skip inserted

%%%%%%%%%%%%%%%%%%%%%%%%%%%%%%%
%%%%%%%%%%%%%%%%%%%%%%%%%%%%%%%
%%%%%%%%%%%%%%%%%%%%%%%%%%%%%%%

\end{document}